\documentclass[12pt]{article}
\usepackage{epsfig,graphicx}
\usepackage{fpcp03}

\def\beq{\begin{equation}}
\def\eeq#1{\label{#1}\end{equation}}
\def\eeqn{\end{equation}}

\def\beqa{\begin{eqnarray}}
\def\eeqa#1{\label{#1}\end{eqnarray}}
\def\eeqan{\end{eqnarray}}

\def\Journal#1#2#3#4{{#1} {\bf #2}, #3 (#4)}

\let\bar=\overbar

\def\Dslash{\not{\hbox{\kern-4pt $D$}}}
\def\dslash{\not{\hbox{\kern-2pt $\del$}}}

\def\msb{{\bar{\ssstyle M \kern -1pt S}}}

\def\BB0bar{B^0 {\overline B}^0}
\def\BB0dbar{B_d^0 {\overline B}_d^0}
\def\BB0sbar{B_s^0 {\overline B}_s^0}

\RequirePackage{xspace}

\usepackage{relsize}
\def\babar{\mbox{\slshape B\kern-0.1em{\smaller A}\kern-0.1em
    B\kern-0.1em{\smaller A\kern-0.2em R}}}

\def\b     {\ensuremath{b}\xspace}

\def\Kbar  {\kern 0.2em\overline{\kern -0.2em K}{}\xspace}

\def\Kz    {\ensuremath{K^0}\xspace}
\def\Kzb   {\ensuremath{\Kbar^0}\xspace}
\def\KzKzb {\ensuremath{\Kz \kern -0.16em \Kzb}\xspace}
\def\Kp    {\ensuremath{K^+}\xspace}
\def\Km    {\ensuremath{K^-}\xspace}

\def\KpKm  {\ensuremath{\Kp \kern -0.16em \Km}\xspace}

\def\Dbar    {\kern 0.2em\overline{\kern -0.2em D}{}\xspace}

\def\Dz      {\ensuremath{D^0}\xspace}
\def\Dzb     {\ensuremath{\Dbar^0}\xspace}
\def\DzDzb   {\ensuremath{\Dz {\kern -0.16em \Dzb}}\xspace}
\def\Dp      {\ensuremath{D^+}\xspace}
\def\Dm      {\ensuremath{D^-}\xspace}

\def\DpDm    {\ensuremath{\Dp {\kern -0.16em \Dm}}\xspace}

\def\Bbar    {\kern 0.18em\overline{\kern -0.18em B}{}\xspace}

\def\BB      {\ensuremath{B\Bbar}\xspace}
\def\Bz      {\ensuremath{B^0}\xspace}
\def\Bzb     {\ensuremath{\Bbar^0}\xspace}
\def\BzBzb   {\ensuremath{\Bz {\kern -0.16em \Bzb}}\xspace}
\def\Bu      {\ensuremath{B^+}\xspace}
\def\Bub     {\ensuremath{B^-}\xspace}

\def\BpBm    {\ensuremath{\Bu {\kern -0.16em \Bub}}\xspace}

\mathchardef\Upsilon="7107
\def\Y#1S{\ensuremath{\Upsilon{(#1S)}}\xspace}

\mathchardef\Deltares="7101
\mathchardef\Xi="7104
\mathchardef\Lambda="7103
\mathchardef\Sigma="7106
\mathchardef\Omega="710A

\def\Deltabar{\kern 0.25em\overline{\kern -0.25em \Deltares}{}\xspace}
\def\Lbar{\kern 0.2em\overline{\kern -0.2em\Lambda\kern 0.05em}\kern-0.05em{}\xspace}
\def\Sigbar{\kern 0.2em\overline{\kern -0.2em \Sigma}{}\xspace}
\def\Xibar{\kern 0.2em\overline{\kern -0.2em \Xi}{}\xspace}
\def\Obar{\kern 0.2em\overline{\kern -0.2em \Omega}{}\xspace}
\def\Nbar{\kern 0.2em\overline{\kern -0.2em N}{}\xspace}
\def\Xb{\kern 0.2em\overline{\kern -0.2em X}{}\xspace}

\newcommand{\tev}{\ensuremath{\mathrm{\,Te\kern -0.1em V}}\xspace}
\newcommand{\gev}{\ensuremath{\mathrm{\,Ge\kern -0.1em V}}\xspace}
\newcommand{\mev}{\ensuremath{\mathrm{\,Me\kern -0.1em V}}\xspace}
\newcommand{\kev}{\ensuremath{\mathrm{\,ke\kern -0.1em V}}\xspace}
\newcommand{\ev}{\ensuremath{\mathrm{\,e\kern -0.1em V}}\xspace}
\newcommand{\gevc}{\ensuremath{{\mathrm{\,Ge\kern -0.1em V\!/}c}}\xspace}
\newcommand{\mevc}{\ensuremath{{\mathrm{\,Me\kern -0.1em V\!/}c}}\xspace}
\newcommand{\gevcc}{\ensuremath{{\mathrm{\,Ge\kern -0.1em V\!/}c^2}}\xspace}
\newcommand{\mevcc}{\ensuremath{{\mathrm{\,Me\kern -0.1em V\!/}c^2}}\xspace}

\def\mus  {\ensuremath{\rm \,\mus}\xspace}

\def\mus        {\ensuremath{\,\mu{\rm s}}\xspace}    

\def\to                 {\ensuremath{\rightarrow}\xspace}

\def\pep2{PEP-II}

\def\gsim{{~\raise.15em\hbox{$>$}\kern-.85em
          \lower.35em\hbox{$\sim$}~}\xspace}
\def\lsim{{~\raise.15em\hbox{$<$}\kern-.85em
          \lower.35em\hbox{$\sim$}~}\xspace}

\xspace

\newcommand{\epj}       [1]  {\epjBase\ {\bf #1}}

\def\jetset74   {\mbox{\tt Jetset \hspace{-0.5em}7.\hspace{-0.2em}4}\xspace}

\def\dec{\rightarrow}
\def\ups{$\Upsilon$(4S)}
\def\vub{$V_{ub}$}
\def\vcb{$V_{cb}$}
\def\Journal#1#2#3#4{#1 {\bf #3}, #4 (#2)}

\def\epj{{ Eur. Phys. J.} C}
\begin{document}


\Title{B meson semileptonic decays}
\bigskip


%
\label{ArtusoStart}

%
\author{ Marina Artuso\index{Artuso, M.} }

%
\address{Department of Physics\\
Syracuse University \\
Syracuse, NY 13244, USA \\
}
 \makeauthor\abstracts{ B meson semileptonic decays
are a crucial tool in our studies of the quark mixing parameters
\vcb\ and \vub . The interplay between experimental and
theoretical challenges to achieve precision in the determination
of these fundamental parameters is discussed.}

\centerline{\em Invited talk given at FPCP 2003, Paris, June 3-6,
2003}

\section{Introduction}

In the framework of the Standard Model, the quark sector is
characterized by a rich pattern of flavor-changing transitions,
described by the Cabibbo-Kobayashi-Maskawa (CKM) matrix:
\begin{equation}
V_{CKM} =\left(\begin{array}{ccc}
V_{ud} &  V_{us} & V_{ub} \\
V_{cd} &  V_{cs} & V_{cb} \\
V_{td} &  V_{ts} & V_{tb} \end{array}\right).
\end{equation}
Since the CKM matrix must be unitary, it can be expressed as a
function of only four parameters. It is expected that a detailed
experimental study of the quark mixings may lead to uncover
discrepancies and may define the path towards a more satisfactory
understanding of flavor.

$B$ meson semileptonic decays allow the determination of the
magnitudes of the quark mixing parameters $| V_{cb}| $ and $|
V_{ub}|$, discussed by M. Calvi \cite{calvi} and E. Thorndike
\cite{thorn}. The goal of precision measurements of these
quantities is strongly affected by our progress in evaluating
hadronic matrix elements. While theorists develop calculation
tools of increasing sophistication, the experimental program can
pin down important theoretical parameters and perform crucial
tests of the theory. I will focus my discussion on how we can use
experimental studies of $B$ meson semileptonic decays as probes of
strong interaction dynamics.

\section{The inclusive decay $B\rightarrow X_c \ell \bar{\nu}$.}

  Three decades of experimental
and theoretical studies have lead to considerable progress in our
understanding of the Cabibbo favored inclusive decay $B\rightarrow
X_c \ell \bar{\nu}$. The Operator Product Expansion (OPE) yields
the heavy quark inclusive decay rates as an asymptotic series in
inverse powers of the heavy quark mass. In actuality, several mass
scales are relevant: the b quark mass $m_b$, the c quark mass
$m_c$ and the energy release $E_r\equiv m_b -m_c$
\cite{kolya-2001}. The uncertainties in the predicted $\Gamma
_{sl}/V_{cb}$ have been discussed in numerous theoretical papers
\cite{benson-et-al-durham03}. In parallel, it is important to
probe the theory with a variety of approaches that allow to test
the consistency of the overall picture.

The key parameter in the theoretical expression for the
semileptonic width is $m_b$. As the bare quark mass is affected by
perturbative and non-perturbative contributions, considerable
attention has been devoted to its proper definition \cite{luke02},
\cite{hoang}. Similarly, $m_c$ is a parameter in the hadronic
matrix element and, recently, it has been argued that an
independent experimental determination of this parameter is
opportune \cite{benson-et-al-durham03}. Previous determinations of
the OPE parameters have relied on the value of the difference
$(m_b-m_c)$ obtained the spin averaged meson mass difference
$(\bar{M_B}- \bar{M_D})$ \cite{falk}.

 The leading
non-perturbative corrections arise only to order $1/m_b^2$ and are
parameterized by the quantities $\mu_{\pi}^2\ ({\rm or}\ -\lambda
_1)$ \cite{falk}, \cite{kolya} related to the expectation value of
the kinetic energy of the $b$ quark inside the $b$ hadron, and
$\mu _G^2\ ({\rm or}\ \lambda _2)$ \cite{falk}, \cite{kolya}
related to the expectation value of the chromomagnetic operator.
Quark-hadron duality is an important {\it ab initio} assumption in
these calculations. While several authors \cite{bigiduality} argue
that this ansatz does not introduce appreciable errors as they
expect that duality violations affect the semileptonic width only
in high powers of the non-perturbative expansion, other authors
recognize that an unknown correction may be associated with this
assumption \cite{buchalla}. Arguments supporting a possible
sizeable source of errors related to the assumption of
quark-hadron duality have been proposed \cite{nathan}.  Most of
the experimental studies have focused on the lepton energy and the
invariant mass $M_X$ of the hadronic system recoiling against the
lepton-$\bar{\nu}$ pair.

I will start the discussion with the experimental studies of the
moments of inclusive distributions. CLEO published the first
measurement of the moments of the $M_X^2$ distributions. This
analysis includes a 1.5 GeV/c momentum cut, that allows them to
single out the desired $b \rightarrow c \ell^- \bar{\nu}$ signal
from the ``cascade" $b\rightarrow c \rightarrow s \ell ^+\nu$
background process. The hermeticity of the CLEO detector is
exploited to reconstruct the $\nu$ 4-momentum vector. Moreover,
the $B\bar{B}$ pair is produced nearly at rest and thus it allows
a determination of $M_X$ from the $\nu$ and $\ell$ momenta. They
obtain $<M_X^2-\bar{M_D}^2> = 0.251 \pm 0.066\ {\rm GeV}^2$ and
$<(M_X^2-\bar{M_D}^2>)^2> = 0.576 \pm 0.170 \ {\rm GeV}^4$, where
$\bar{M_D}$ is the spin-averaged mass of the $D$ and $D^{\star}$
mesons. The lepton momentum cut may reduce the accuracy of the OPE
predictions, because restricting the kinematic domain may
introduce quark-hadron duality violations. The shape of the lepton
spectrum provides further constraints on OPE. Moments of the
lepton momentum with a cut $p_{\ell}^{CM}\ge 1.5$ GeV/c have been
measured by the CLEO collaboration \cite{prd2003}. The two
approaches give consistent results, although the technique used to
extract the OPE parameters has still relatively large
uncertainties associated with the $1/m_b^3$ form factors. The
sensitivity to $1/m_b^3$ corrections depends upon which moments
are considered. Bauer and Trott \cite{bt-2001} have performed an
extensive study of the sensitivity of lepton energy moments to
non-perturbative effects. In particular, they have proposed
``duality moments," very insensitive to neglected higher order
terms. The comparison between the CLEO measurement of these
moments\cite{prd2003} and the predicted values shows a very
impressive agreement:
\begin{eqnarray}
D_3&\equiv& {\int_{1.6\,{\rm GeV}} E_\ell^{0.7}{d\Gamma\over
dE_\ell}\,dE_\ell\over \int_{1.5\,{\rm GeV}}
E_\ell^{1.5}{d\Gamma\over dE_\ell}\,dE_\ell}=\cases{0.5190\pm
0.0007&(T)\cr
0.5193\pm 0.0008&(E)} \nonumber\\
D_4&\equiv& {\int_{1.6\,{\rm GeV}} E_\ell^{2.3}{d\Gamma\over
dE_\ell}\,dE_\ell\over \int_{1.5\,{\rm GeV}}
E_\ell^{2.9}{d\Gamma\over dE_\ell}\,dE_\ell}=\cases{0.6034\pm
0.0008&(T)\cr
0.6036\pm 0.0006&(E)} \nonumber\\
\end{eqnarray}
(where ``T" and ``E" denote theory and experiment, respectively).

\begin{figure}[htb]
\begin{center}
\epsfig{file=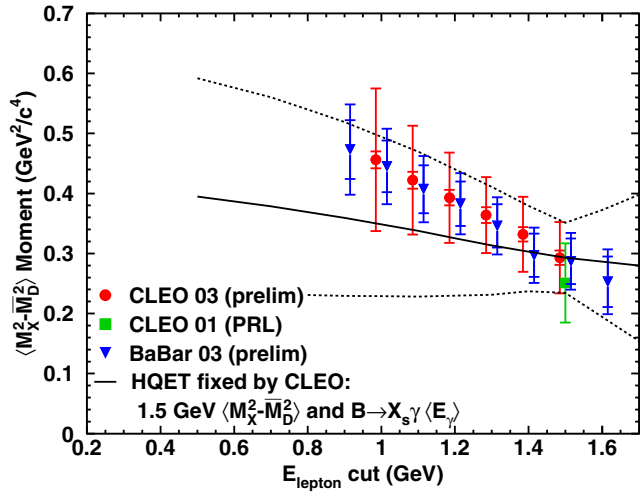,height=70mm} \caption{The results of
the recent CLEO analysis \cite{newmxmom} compared to previous
measurements \cite{ref:oldhadmom,ref:babarmoms} and the HQET
prediction. The theory bands shown in the figure reflect the
variation of the experimental errors on the two constraints, the
variation of the third-order HQET parameters by the scale $(0.5\
{\rm GeV})^3$, and variation of the size of the higher order QCD
radiative corrections \cite{bauer-et-al-2002}.}
\label{mom-leptons}
\end{center}
\end{figure}

More recently, both CLEO and BaBar explored the moments of the
hadronic mass $M_X^2$ with lower lepton momentum cuts. In order to
identify the desired semileptonic decay from background processes
including cascade decays, continuum leptons and fake leptons, CLEO
performs a fit for the contributions of signal and backgrounds to
the full three-dimensional differential decay rate distribution as
a function of the reconstructed quantities $q^2$, $M_X^2$,
$\cos{\theta _{W\ell}}$. The signal includes the components
$B\rightarrow D \ell \bar{\nu}$, $B\rightarrow D^{\star} \ell
\bar{\nu}$, $B\rightarrow D^{\star\star} \ell \bar{\nu}$,
$B\rightarrow X_c \ell \bar{\nu}$ non-resonant and $B\rightarrow
X_u \ell \bar{\nu}$. The backgrounds considered are: secondary
leptons, continuum leptons and fake leptons.
Fig.~\ref{mom-leptons} shows the extracted $<M_X^2-\bar{M}_D^2>$
moments as a function of the minimum lepton momentum cut from
these two measurements, as well as the original measurement with
$p_{\ell}\ge 1.5$ GeV/c. The results are compared with theory
bands that reflect experimental errors, $1/m_b^3$ correction
uncertainties and uncertainties in the higher order QCD radiative
corrections \cite{bauer-et-al-2002}. The CLEO and BaBar results
are consistent and show an improved agreement with theoretical
predictions with respect to earlier preliminary results
\cite{babar-ichep02}. Moments of the $M_X$ distribution without an
explicit lepton momentum cut have been extracted from preliminary
DELPHI data \cite{battaglia} and give consistent results.

\begin{figure}[htb]
\begin{center}
\epsfig{file=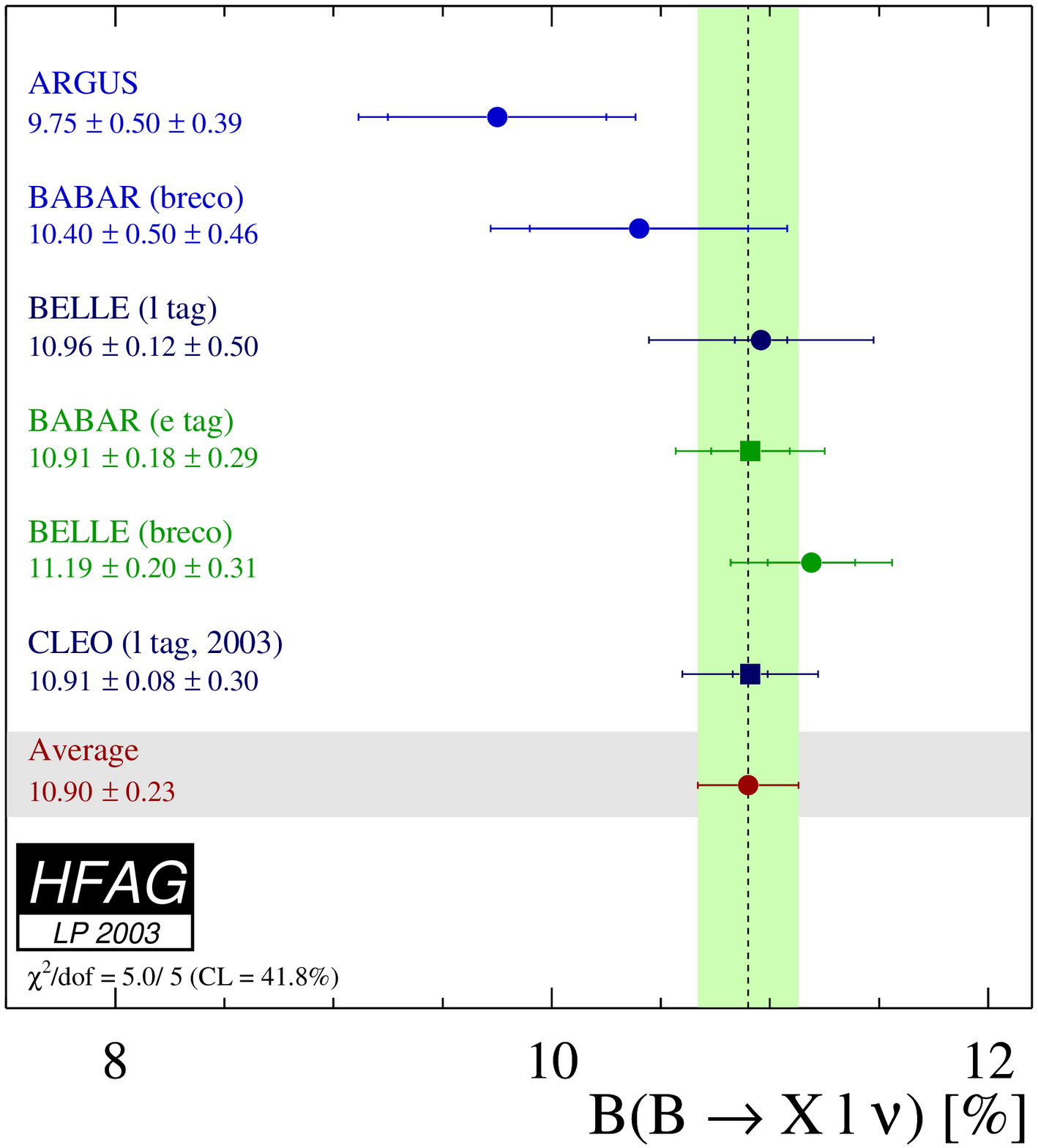,height=80mm} \caption{Summary of
model independent semileptonic branching fraction measurements
performed at the $\Upsilon(4S)$ center-of-mass energy.}
\label{sl-inclusive-4s}
\end{center}
\end{figure}

Experiments operating at the \ups\ center-of-mass energy use a
dilepton sample to separate the decay process $b\rightarrow c
\ell^- \bar{\nu}$ (primary leptons) from the $b\rightarrow
c\rightarrow s \ell ^+\nu$ (cascade leptons). This technique
allows a direct determination of the primary lepton spectrum over
almost all the range kinematically allowed. Thus the semileptonic
branching fraction extracted from this measurement has almost no
model dependence.  Fig.~\ref{sl-inclusive-4s} shows a summary of
the \ups\ measurements of inclusive semileptonic branching
fractions. The overall experimental error is of the order of 2\%,
quite impressive given the number of systematic effects that need
to be addressed in this measurement.

At LEP, ${\cal B}(\rm b \to X \ell \nu)$ is measured with
dedicated analyses \cite{aleph}, \cite{D}, \cite{L}, \cite{O},
summarized in (Table~\ref{t:inc}). The average LEP value for
${\cal B} (\rm b \to X \ell \nu) = (10.59 \pm 0.09 \pm 0.30)\% $
is taken from a fit \cite{EWWG}, \cite{vcb-pdg2002} which combines
the semileptonic branching ratios, $\rm B^0-\bar{B^0}$ mixing, and
$\rm R_b =\Gamma(Z\to b\bar{b})/\Gamma(Z\to had)$. In this
average, Artuso and Barberio use the modelling error quoted by
\cite{aleph} rather than the error from the combined fit, as the
ALEPH procedure is based on more recent information. The dominant
errors in the combined branching fraction are the modelling of
semileptonic decays (2.6\%) and the detector related items
(1.3\%).

\begin{table}[hbtp]
\begin{center}
\begin{tabular}{|l|c|} \hline
 Experiment &  ${\cal B}(b\dec\ell\nu) \%$  \\
 \hline
 ALEPH      &  10.70 $\pm$ 0.10 $\pm$ 0.23 $\pm$ 0.26 \\
 DELPHI     &  10.70 $\pm$ 0.08 $\pm$ 0.21 $\pm^{+0.44}_{-0.30}$ \\
 L3         &  10.85 $\pm$ 0.12 $\pm$ 0.38 $\pm$ 0.26 \\
 L3 (double-tag)         &  10.16 $\pm$ 0.13 $\pm$ 0.20 $\pm$ 0.22 \\
 OPAL       &  10.83 $\pm$ 0.10 $\pm$ 0.20 ${\pm}^{+0.20}_{-0.13}$ \\ \hline
 LEP Average & 10.59 $\pm$ 0.09 $\pm$ 0.15 $\pm$ 0.26 \\ \hline
\end{tabular}
\end{center}
\caption{$ {\cal B} (\rm b \to \ell )$ measurement from LEP and
their average. The errors quoted reflect statistical, systematic
and modelling uncertainties respectively. \label{t:inc}}
\end{table}

\section{Exclusive $B$ charmed semileptonic decays}

The exclusive decay channels $B\dec D^{\star}\ell \bar{\nu}$ and
$B\dec D\ell \bar{\nu}$ have been studied extensively because they
provide an alternative avenue to extract the CKM parameter
$V_{cb}$. In this case, the theoretical tool to extract the
hadronic matrix element is Heavy Quark Effective theory (HQET)
\cite{isgur-wise}. HQET relates the differential partial decay
width $d\Gamma/dw$, where $w$ is the inner product of the $B$ and
$D^{\star}$ meson 4-velocities.

In particular, $d\Gamma/dw$ for the decay $B\dec D^{\star}\ell
\bar{\nu}$ is given by:
\begin{equation}
\frac{d\Gamma}{dw}(B\rightarrow D^{\star}\ell \nu) = \frac{G_F^2
|V_{cb}|^2}{48\pi^3}{\cal K}(w){\cal F}(w)^2,
\end{equation}
where ${\cal K}(w)$ is a known phase space factor and the form
factor ${\cal F}(w)$ is generally expressed as the product of a
normalization factor ${\cal F}(1)$ and a function, $g(w)$,
constrained by dispersion relations \cite{grinstein}. By virtue of
Luke's theorem \cite{luke}, the first term in the non-perturbative
expansion in powers of $1/m_Q$ vanishes. This property, as well as
the large branching fraction, make this decay the central focus of
the experimental determination of $V_{cb}$ from exclusive
semileptonic decays in recent years. It has been studied both at
the $\Upsilon(4S)$ and at the $Z^0$ resonance center-of-mass
energy at LEP.

The $\Upsilon(4S)$ has the advantage that the $B\bar{B}$ pair is
produced nearly at rest, so that kinematical constraints can be
used to determine the $\nu$ momentum and separate the signal from
the dominant sources of background. Yields are obtained as a
function of the variable $w$. The LEP experiments have a uniform
efficiency as a function of the lepton momentum because of the
higher boost at their center-of-mass energy. However, there is no
simple kinematic tool to separate signals from background. In
particular, contributions from higher mass charmed final states
are evaluated using models that apply only to narrow resonances.
Thus a detailed knowledge of the yet unknown broad resonances and
of non-resonant $D^{(\star)} \rm n \pi$ production have to be
acquired to provide reliable systematic errors in this method.

New interest on the decay $B\dec D \ell \bar{\nu}$ has been
spurred by recent developments \cite{lepage} in lattice gauge
calculation techniques that include the decay $B\dec D \ell
\bar{\nu}$ among a selected class of ``golden modes" whose
hadronic matrix element will be calculated with high precision.
This decay mode is more challenging to experimental study because
of its smaller branching fraction. Nonetheless both the branching
fraction and differential decay distributions such as $d\Gamma/dw$
have been measured by several experiments.

Table~\ref{charm-excl} summarizes our present knowledge on the
branching fractions to the ``ground state" decay modes $D\ell
\bar{\nu}$ and $D^{\star}\ell\bar{\nu}$.  One interesting question
is whether the composition of the total semileptonic width is
consistent with our theoretical understanding of semileptonic
decays. In the Shiftman-Voloshin limit, the total semileptonic
width is expected to be saturated by these ground state decay
modes. All other states contribute to the decay width only to
order $1/m_b^2$ because of heavy quark symmetry. However, if we
consider the present world averages for the inclusive and
exclusive decay modes, we note that about 30\% of the semileptonic
width includes orbitally excited final states or higher mass
objects. A more detailed knowledge of the final states and the
dynamics involved in this poorly known component of the
semileptonic decay rate is very important to assess our
understanding of $b$ meson semileptonic decays. In particular, a
study of the differential decay rate d$\Gamma$/d$M_X$ in the $M_X$
mass region about 3 GeV \cite{vademecum} have been proposed as a
check of quark-hadron duality violations. In fact, earlier
predictions of possible sizeable duality violation \cite{nathan}
have also indicated high mass tails as the possible manifestation
of such effects.

\begin{table}[htbp]
\begin{center}
\begin{tabular}{|c|l|l|}
\hline
Experiment & ${\cal B}(\bar{B}\dec D^{\star}\ell \bar{\nu})(\%) $ & Remarks\\
\hline
CLEO\cite{cleo-ds} & $6.25\pm 0.19 \pm 0.39$& both $D^{\star +}$ and $D^{\star o}$ final states\\
BaBar\cite{babar-ds} & $4.68\pm 0.03 \pm 0.28$ & preliminary \\
Belle\cite{belle-ds} & $4.78\pm 0.23 \pm 0.43$ & ~~\\
ALEPH\cite{aleph-ds} & $5.75\pm 0.26 \pm 0.36$ & \\
OPAL \cite{opal-ds}& $5.44\pm 0.19 \pm 0.41$ & Full reconstruction \\
OPAL \cite{opal-ds}& $6.15\pm 0.27 \pm 0.58$ & Partial reconstruction \\
DELPHI\cite{del-p-ds} & $5.02\pm 0.13 \pm 0.36$ & Partial reconstruction \\
DELPHI\cite{del-p-ds} & $5.70\pm 0.20 \pm 0.41$ & Full reconstruction (prelim)\\
\hline Average & $5.20\pm 0.19$ & HFAG Summer 03 Average\\
\hline \hline
Experiment & ${\cal B}(\bar{B}\dec D\ell \bar{\nu})(\%)$ & ~~\\
\hline
CLEO \cite{cleo-d} & $2.11\pm 0.13\pm 0.17$ &  \\
Belle \cite{belle-d} & $2.11\pm 0.13\pm 0.17$ & \\
ALEPH \cite{aleph-ds}& $2.42\pm 0.20\pm 0.50$& \\
\hline
Average & $2.14\pm 0.20$&HFAG Summer 03 Average\\
\hline
\end{tabular}
\caption{Exclusive branching fractions for ${\cal B}(B\dec
D^{(\star)}\ell \bar{\nu})$ rescaled to common inputs by the Heavy
Flavor Averaging Group (HFAG) \cite{hfag}} \label{charm-excl}
\end{center}
\end{table}

\section{The inclusive charmless semileptonic decay $B\dec X_u \ell
\bar{\nu}$}

Charmless $B$ meson semileptonic decays constitute only about 1\%
of the total semileptonic width. Thus the big experimental
challenge is the high suppression needed for this dominant
background. Typical strategies involve severe cuts on the phase
space that destroy the convergence of the OPE expansion
\cite{luke-durham}. Two approaches have been taken: CLEO, ARGUS,
and, more recently, BaBar and Belle have used a lepton momentum
cut to minimize the charm background. Alternatively, the invariant
mass  $M_X$ of the hadron system recoiling against the
lepton-$\bar{\nu}$ pair, has been used to identify the charmless
component, typically involving lighter $M_X$. Both these
approaches involve kinematic regions where the differential decay
rate is very sensitive to the details of the wave function of the
$b$ quark inside the $B$ hadron \cite{luke16}. One can reduce this
theoretical uncertainty by considering more complex kinematic cuts
\cite{ligeti-luke}, or by measuring the universal structure
function, which describes the distribution of the light-cone
component of the residual momentum of the $b$ quark in some other
processes \cite{luke16}, \cite{luke23}.

CLEO \cite{cleo-improved} uses the shape of the $\gamma$ spectrum
in $B \dec X_s \gamma$ to determine the quantity $f_u(p)$ giving
the fraction of the lepton spectrum in $B\dec X_u \ell \bar{\nu}$
comprised within a given momentum cut. Table~\ref{bu-incl}
summarizes the CLEO results as well as a similar measurement
performed by BaBar. They restrict their measurement to the lepton
interval 2.3-2.6 GeV and use the $f_u(2.3\ {\rm GeV})$ measured by
CLEO to extrapolate to the whole momentum interval. Note that the
lower the momentum cut the larger the uncertainty induced by
subleading corrections and by weak annihilation graphs
\cite{luke-durham}. Hopefully future measurements will involve
larger momentum intervals. Moreover, a precise comparison between
$D$ and $D_S$ semileptonic decays will enable us to determine weak
annihilation effects more precisely \cite{voloshin}.

\begin{table}[htbp]
\begin{center}
\begin{tabular}{|c|l|l|}
\hline
Experiment & ${\cal B}(\bar{B}\dec X_u \ell \bar{\nu})(\times 10^3)$ & Remarks\\
\hline
CLEO \cite{cleo-improved} & $1.77\pm 0.29_{exp}\pm 0.38_{f_u}$ & $p_{\ell}=2.2-2.6$ GeV/c \\
BaBar \cite{babar-bu} & $2.05\pm 0.27_{exp}\pm 0.46_{f_u}$&$p_{\ell}=2.3-2.6$ GeV/c +CLEO $f_u$\\
\hline
Babar\cite{delre}  & $2.24 \pm 0.27_{stat}\pm 0.26_{sys}\pm 0.39_{th}$& ($M_X$)\\
Belle\cite{sugiyama} & $2.62 \pm 0.63_{stat}\pm 0.23_{sys}\pm 0.05_{b\dec c}\pm 0.41_{\b\dec u}$ & Prelim. ($M_X$)\\
Belle \cite{sugiyama}& $1.64 \pm 0.14_{stat}\pm 0.36_{sys}\pm
0.28_{b\dec c}\pm
0.22_{\b\dec u}$ & Prelim. - Advanced $\nu$ reco.\\
 \hline
\end{tabular}
\caption{Inclusive ${\cal B}(\bar{B}\dec X_u \ell \bar{\nu})$
data.} \label{bu-incl}
\end{center}
\end{table}

\begin{figure}[htb]
\begin{center}
\epsfig{file=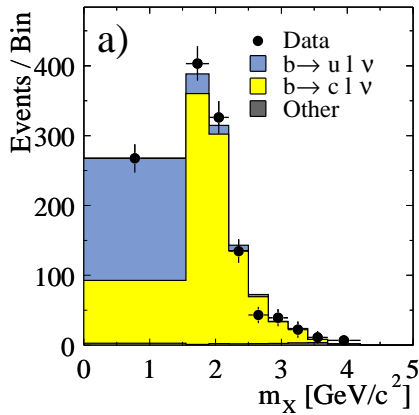,height=60mm} \caption{A $\chi ^2$ fit
to the $M_X$ distribution for the $\bar{B}\dec X_u\ell \bar{\nu} $
enriched sample: shown are the data (points) and the 3 fitted
distribution (BaBar)\cite{delre}.} \label{babar-mass}
\end{center}
\end{figure}

\begin{figure}[htb]
\begin{center}
\epsfig{file=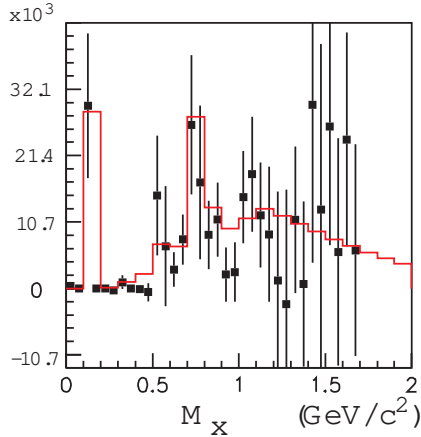,height=60mm} \caption{Efficiency
corrected $M_X$ distribution for data (solid square with errors)
and for a hybrid Monte Carlo model (histogram) (Prelim. Belle)
\cite{sugiyama}.} \label{belle-mass}
\end{center}
\end{figure}

 BaBar and Belle have also explored also studies
of inclusive charmless semileptonic decays using the invariant
mass of the hadronic system $M_X$ recoiling against the
lepton-$\nu$ pair as a discriminant of the $b\dec u$ transition.
These collaborations have started exploiting the high statistics
data samples that they are collecting by performing these studies
with tagging $B$'s. BaBar uses a sample of $B\dec \bar{D} Y$
events, where $Y$ denotes a collection of hadrons with total
charge $\pm 1$ composed of
$n_1\pi^{\pm}+n_2K^{\pm}+n_3K_S+n_4\pi^0$, where $n_1+n_2<6$,
$n_3<3$, and $n_4<3$. They consider leptons with minimum momentum
in the $B$ rest frame $p_{\ell}^\star >1$ GeV/c.
Fig.~\ref{babar-mass} shows the results of a $\chi ^2$ fit to the
$M_X$ distribution for the $\bar{B}\dec X_u \ell \bar{\nu}$
enriched sample. Note that the complex set of cuts considered
\cite{delre} still allows a significant $b\dec c \ell \bar{\nu}$
background component. Belle \cite{sugiyama} uses $B\dec
D^{(\star)} \ell \bar{\nu}$ as a tagging sample, relying on the
well known features and relatively large branching fractions of
these decays. Fig.~\ref{belle-mass} shows preliminary results from
this approach. In addition, Belle is pursuing a more complex $\nu$
reconstruction through an ``annealing" technique to sort decay
products into tagging and $b\dec u \ell \bar{\nu}$. Preliminary
results have been reported \cite{sugi9} and are included in
Table~\ref{bu-incl}. Older measurements from LEP have used the
invariant mass $M_X$ of the hadronic system produced as a
discriminant of charmless semileptonic transition. The background
from $b\dec c$ transitions was significant: the average branching
fraction for $b\dec u \ell\bar{\nu}$ from all the LEP experiments
was found to be $(1.74 \pm 0.37_{exp}\pm 0.38_{b\dec c} \pm
0.21_{b\dec u})\times 10^{-3}$ \cite{lep-vub}.


\section{Exclusive $B$ charmless semileptonic decays}
A considerable body of experimental and theoretical work has been
focused on the final states involving the lightest hadrons in the
final states, namely $\pi$, $\eta$ and $\rho/\omega)$. In this
case, the weak decay involves a heavy to light quark transition,
thus HQET does not apply directly. It has been invoked to relate
heavy to light transitions in the charm and beauty sector
\cite{nathan-heavy-light}. This strategy invokes relationships
between form factors in charm semileptonic decays such as $D\dec
\pi \ell \nu$ to the corresponding ones in $B\dec \pi \ell
\bar{\nu}$ decays. A precise measurement of the former would shed
some light on the latter. Alternatively, lattice gauge
calculations predict the full QCD matrix element calculation.
Indeed, the decay $B\dec \pi \ell \bar{\nu}$ is one of the of the
``golden modes" for which a precision lattice gauge calculation is
planned in the next few years \cite{lepage}. The $B\dec \rho \ell
\bar{\nu}$ poses the additional challenge of involving an unstable
hadron in the final state. This is because unstable hadrons are
strongly affected by finite lattice volume, e.g. the $\pi$'s in
fluctuations $\rho\dec \pi\pi \dec \rho$ can be on-shell and
propagate freely to lattice boundary \cite{lepage-03}. In absence
of rigorous evaluation of the full matrix element, a variety of
models have been used to study these decays, from quark models
\cite{isgw2}, \cite{wbs}, \cite{ball}, to unquenched lattice
calculations \cite{ukqcd}. These different approaches have been
very helpful in understanding  key dynamical features of these
decays, but they all suffer from the difficulty of assessing their
uncertainty in a manner that can systematically be improved (e.g.
with an asymptotic series). There is a considerable spread in the
predictions, as different ansatzs lead to different differential
distributions in the Dalitz plot. In particular, the $q^2$
differential distribution has considerable discriminant power
\cite{ma-qsq}.

\begin{figure}[htb]
\begin{center}
\epsfig{file=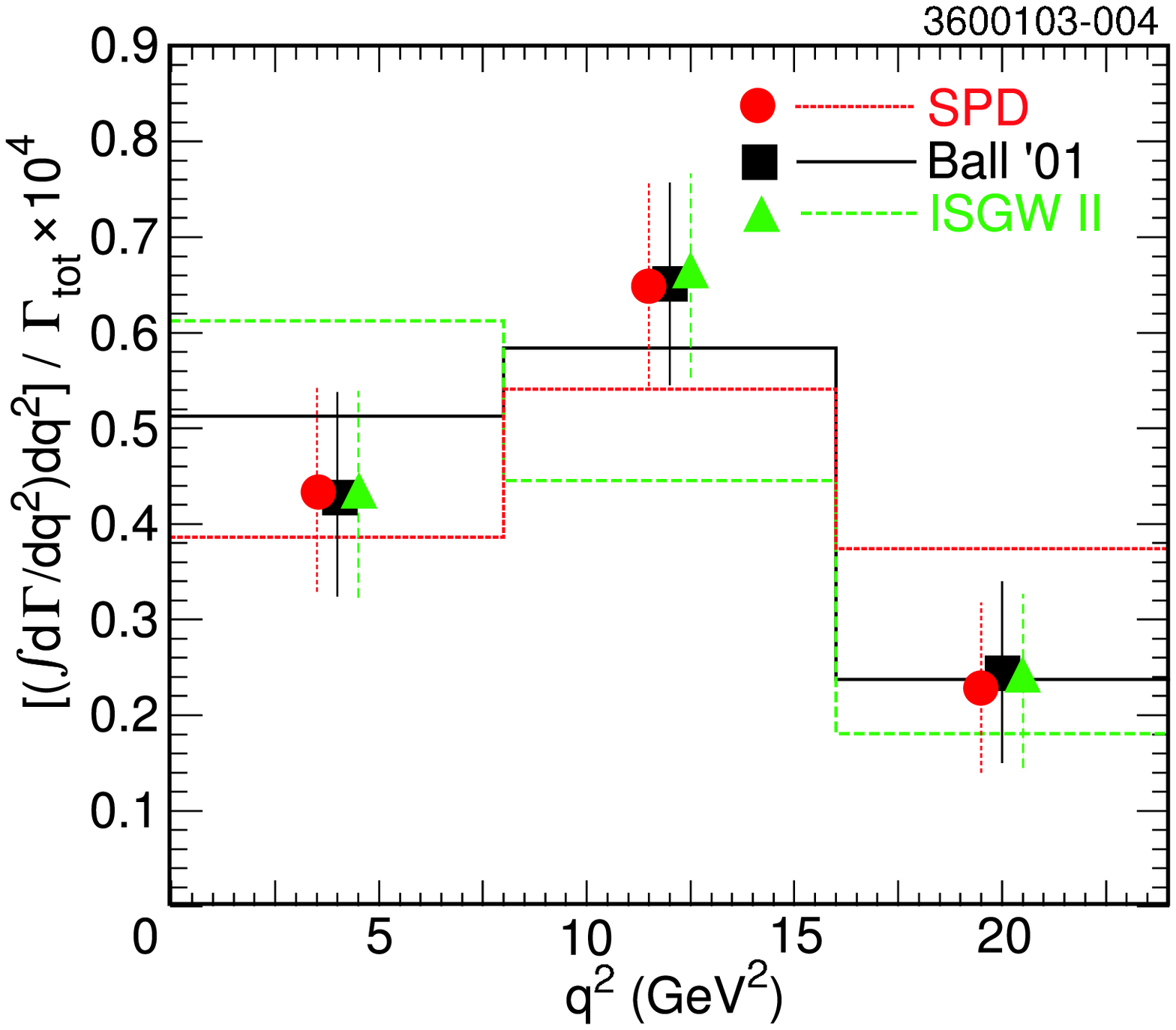,height=50mm}
\epsfig{file=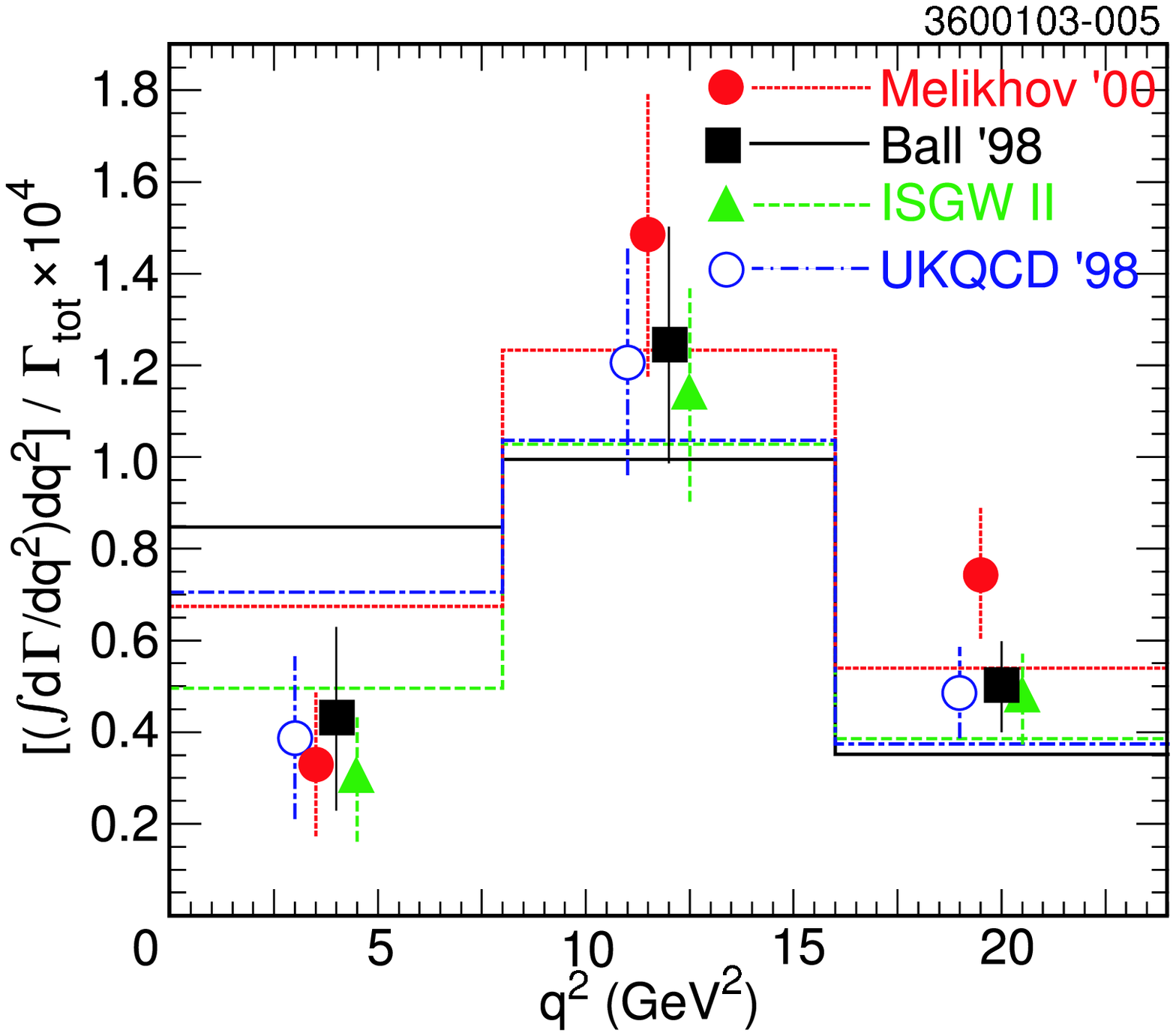,height=50mm}\caption{Measured
branching fractions in the restricted $q^2$ intervals for
$B^0\to\pi^-\ell^+\nu$ (left) and $B^0\to\rho^-\ell^+\nu$ (right).
In addition to the data points, the best fits to the predicted
$d\Gamma/dq^2$ (histograms) for the three models used to extract
both rates and $|V_{ub}|$ are shown as histograms. The data points
have small horizontal offsets introduced for clarity. The last bin
has been artificially truncated at 24 GeV$^2$ in the plot -- the
information out to $q^2_{\rm max}$ has been included in the work
\cite{new-cleo-pirho}.} \label{qsq-cleo}
\end{center}
\end{figure}

From the experimental point of view, there are two important goals
in studying exclusive charmless semileptonic decays. The first is
to lower the minimum lepton momentum cut, to minimize the
extrapolation required to derive the partial width to a specific
final state. The second is to devise analysis methods that
minimize uncertainties arising from the $q^2$ dependence of the
form factors. Great strides have been made in exploiting the
hermeticity of detectors such as CLEO in developing an efficient
$\nu$ reconstruction. This has lead to two important advantages.
On one hand, the phase space of this decay included in the
analyses has become considerably larger. For example, a recent
CLEO result \cite{new-cleo-pirho} uses lepton momenta as low as
1.5 GeV/c. The reconstructed $\nu$ 4-momentum vector allows to
perform a full reconstruction of the decay, thus exploiting the
same analysis methods used in hadronic $B$ decays. One interesting
feature of this analysis is that the branching fractions are
determined independently in three $q^2$ regions.
Fig.~\ref{qsq-cleo} shows the measured branching fractions in the
restricted $q^2$ intervals, as well as the best fit to the
predicted $d\Gamma/dq^2$ in 4 different models. It can be seen
that this comparison already shows some discriminant power, that
will be fully exploited in the future.

\section{Conclusions}

Experimental studies of $B$ meson semileptonic decays have reached
a level of great sophistication, but several issues still need to
be addressed. Progress will involve a close interplay between
experimental and theoretical work. Experimental studies that are
crucial to further progress include:
\begin{itemize}
\item[-] a full understanding of the composition of the $B$ meson
semileptonic width; \item [-] a precise determination of the OPE
expansion parameters; \item [-] overall checks of the consistency
of the theoretical picture, including extent of quark-hadron
duality violation and effects of neglected terms in the
perturbative and non-perturbative expansion;\item[-] study of
inclusive charmless semileptonic decays in selected phase space
domains sensitive to different theoretical effects;\item[-]
detailed studies of the kinematic properties of charmless
semileptonic decays with tagged samples. \end{itemize}

The theoretical challenge is to increase the reliability and the
precision of the evaluation of hadronic matrix elements in a
regime where non-perturbative effects are quite important. An
example of a winning strategy is the lattice community plan to
pursue the calculation of a whole host of "golden modes"
\cite{lepage}. This program is structured as a two prong approach.
Predictions on golden modes in the charm sector will be tested in
precision studies at upcoming charm factories, such as CLEO-c,
that is presently starting its experimental program at the $\psi
^{\prime\prime}$. This validation will be critical to develop the
confidence on the errors in theoretical quantities relative to $b$
decays.

The completion of this complex and ambitious program will lead to
precise measurements of the CKM parameters \vcb\ ad \vub\ and thus
will provide a unique contribution to the multifaceted challenge
to the Standard Model through the study of beauty and charm
decays.

\section{Acknowledgments}
I would like to thank E. Barberio and S. Stone for interesting
discussions. I would also like to thank S. Akira, T. Browder, M.
Calvi, H. Kakuno, U. Langenegger, and A. Limosani for their help
in collecting the information summarized in this paper. This work
was supported from the US National Science Foundation.

\end{document}